\documentclass[iop]{emulateapj}
\usepackage{graphicx}
\usepackage{rotating}

\shorttitle{Dynamics of multi-cored magnetic structures}
\shortauthors{Requerey et al.}

\begin{document}


\title{Dynamics of multi-cored magnetic structures in the quiet Sun}


\author{Iker S. Requerey, Jose Carlos Del Toro Iniesta, and Luis R. Bellot Rubio}
\affil{Instituto de Astrof\'{i}sica de Andaluc\'{i}a (CSIC), Apdo. de Correos 3004,
    E-18080 Granada, Spain}
\email{iker@iaa.es}
\author{Valent\'{i}n Mart\'{i}nez Pillet}
\affil{National Solar Observatory, Boulder, CO 80303, USA}
\affil{Instituto de Astrof\'{i}sica de Canarias, Avda. V\'{i}a L\'{a}ctea s/n, La Laguna, Spain}
\author{Sami K. Solanki}
\affil{Max-Planck Institut f\"{u}r Sonnensystemforschung, Justus-von-Liebig-Weg 3, D-37077, G\"{o}ttingen, Germany}
\affil{School of Space Research, Kyung Hee University, Yongin, 446-701 Gyeonggi, Korea}
\author{Wolfgang Schmidt}
\affil{Kiepenheuer-Institut f\"{u}r Sonnenphysik, Sch\"{o}neckstr. 6, D-79104, Freiburg, Germany}



\begin{abstract}

We report on the dynamical interaction of quiet-Sun magnetic fields and granular convection in the solar photosphere as seen by \textsc{Sunrise}. We use high spatial resolution (0\farcs 15--0\farcs 18) and temporal cadence (33 s) spectropolarimetric Imaging Magnetograph eXperiment data, together with simultaneous CN and Ca\,\textsc{ii}\,H filtergrams from \textsc{Sunrise} Filter Imager. We apply the SIR inversion code to the polarimetric data in order to infer the line of sight velocity and vector magnetic field in the photosphere. The analysis reveals bundles of individual flux tubes evolving as a single entity during the entire 23 minute data set. The group shares a common canopy in the upper photospheric layers, while the individual tubes continually intensify, fragment and merge in the same way that chains of bright points in photometric observations have been reported to do. The evolution of the tube cores are driven by the local granular convection flows. They intensify when they are ``compressed'' by surrounding granules and split when they are ``squeezed'' between two moving granules. The resulting fragments are usually later regrouped in intergranular lanes by the granular flows. The continual intensification, fragmentation and coalescence of flux results in magnetic field oscillations of the global entity. From the observations we conclude that the magnetic field oscillations first reported by \citet{2011ApJ...730L..37M} correspond to the forcing by granular motions and not to characteristic oscillatory modes of thin flux tubes.

\end{abstract}



\section{Introduction}

Most of our empirical knowledge of the structure and dynamics of quiet-Sun magnetism derives from observations of the solar photosphere. In this thin layer, magnetic energy is in many places of the same order as the kinetic energy. Therefore, the interaction between the magnetic field and convection at the solar surface is an efficient way of converting kinetic energy into form that can be transported to the upper layers of the solar atmosphere by the magnetic field.

The most direct method of detecting the solar magnetic field is by measuring polarized light generated via the Zeeman effect. Unfortunately, in the quiet Sun, the Zeeman effect produces only a weak polarization signal, whose measurement requires both,  high spatial resolution and accurate polarimetric sensitivity. Such measurements have only recently been achieved by the \textit{Hinode} spectro-polarimeter \citep{2013SoPh..283..579L} and the Imaging Magnetograph eXperiment \citep[IMaX;][]{2011SoPh..268...57M} aboard the \textsc{Sunrise} balloon-borne solar observatory \citep{2010ApJ...723L.127S,2011SoPh..268....1B,2011SoPh..268..103B,2011SoPh..268...35G}. 

Before the era of space-borne spectropolarimeters, polarimetric observations have been limited by the need for stable seeing conditions to achieve a high spatial resolution. Instead, to maximize spatial and temporal resolution, indirect signatures, or proxies of magnetic structures have been used. In particular, magnetic elements, usually described in terms of flux tubes, have been tentatively identified with bright points (BPs) in photometric observations. Based on a recent comparison between \textsc{Sunrise} observations and MHD simulations, \citet{2014A&A...568A..13R} deduce that all magnetic BPs are associated with kG magnetic flux concentrations.

White-light observations obtained at the Pic du Midi Observatory in the French Pyr\'{e}n\'{e}es, revealed a mean lifetime of 18 minute for facular \citep{1983SoPh...85..113M} and network \citep{1992SoPh..141...27M} BPs. Many BPs become elongated when they are squeezed between two moving or expanding granules \citep{1994A&A...287..982R}. 70\% of these elongation processes end with the fragmentation of the bright structures.

\citet{1984SoPh...94...33M} were the first to observe bright points in the Fraunhofer G band, a CH molecular band-head around 4305 \AA . At these wavelengths BPs exhibit higher contrast than the one they display in the continuum. \citet{1996ApJ...463..365B} studied the dynamics of G-band bright points observed with the 50 cm Swedish Vacuum Solar Telescope \citep{1985ApOpt..24.2558S} on the island of La Palma, Spain. Driven by the evolution of the local granular convection flows, fragmentation and coalescence are two important processes driving the evolution of BPs. BPs also appear to rotate and fold in chains or groups. Periodically, they split into smaller fragments, merge with other BPs, and sometimes fade until they are no longer distinguishable from their surroundings. Nonetheless, \citet{1998ApJ...495..973B} found some BP groups to persist during the entire 70 minute data set. However, the different members of these chains cannot be identified as individual entities for longer than a granule lifetime, i.e., 6-8 minutes. Consistent with this view, \citet{1996ApJ...463..365B} concluded that the canonical picture of stable, isolated flux tube does not agree with observations.

Simultaneous filtegram and magnetogram observations revealed that continuum and line-core BPs \citep{1992Natur.359..307K,1992ApJ...393..782T,1993SoPh..144....1Y} and G-band BPs \citep{1996ApJ...463..365B,2001ApJ...553..449B} appear associated with a magnetic feature. While isolated BPs  have nearly the same size as the associated magnetic element, BP groups appear inside a large magnetic structure that extends beyond the group. For the largest magnetic structures, \citet{2000A&A...359..373M} found several magnetic signal maxima at the location of individual BPs.

If we assume that BPs are the counterparts of magnetic flux tubes, their fragmentation by the perturbation of surrounding granules might indicate that magnetic elements are liable to the interchange, or fluting, instability \citep[e.g.,][Chap. 5]{1973ppp..book.....K}. \citet{1975SoPh...40..291P} and \citet{1975Ap&SS..34..347P} noticed that the interchange instability is indeed an intrinsic property of flux tubes. However, due to the reduced density of the magnetic plasma, \citet{1977MNRAS.179..741M} showed that flux tubes with fluxes greater than about $10^{19}$ Mx, such as  sunspots and pores, can be stabilized by buoyancy thanks to the rapid expansion with height of their field. Small quiet-Sun magnetic structures with fluxes in the range of $10^{16}$--$10^{18}$ Mx \citep[e.g.][]{1995SoPh..160..277W} obviously do not fulfill this criterion. \citet{1984A&A...140..453S} proposed that such features could be stabilized as well if they are surrounded by whirl flows, with a whirl velocity around magnetic features between 2 and 4 km s$^{-1}$ \citep{1993A&A...268..299B}. Observations of whirl flows were first reported by \citet{2008ApJ...687L.131B}. However, their lifetimes are only about 5 minute, on average, as they often do not survive neighboring granules, which have a similar lifetime \citep{2010ApJ...723L.139B}.

Stabilization of magnetic elements by means of the whirl flow mechanism is restricted to cylindrical flux tube geometry. \citet{1993A&A...276..236B} showed that elongated magnetic slabs, or flux sheets, are also flute unstable. He demonstrated that the slabs are most strongly liable to the instability in a layer close to $\tau_{c} = 1$, where fragmentation into single tube-filaments takes place. These filaments, however, lose their identity at lower and upper layers as they merge into a single, stable magnetic slab. Of course, the validity of such idealized computations in the real, highly dynamic, turbulent solar photosphere remains an open question.

In order to shed new light on the physical mechanism behind the dynamic nature of quiet-Sun magnetism, high spatial and temporal resolution is required over a sufficiently long time series, along with accurate polarimetry. Such high-quality observations have only recently been achieved with \textsc{Sunrise}/IMaX. The unprecedented spatial resolution of 0\farcs 15--0\farcs 18, allowed for the first time photospheric magnetic elements to be spatially resolved even in the quiet Sun internetwork without requiring an ad-hoc filling factor, that specifies the fraction of the pixel filled with magnetic field \citep{2010ApJ...723L.164L}. This represents a considerable advance compared to previous works that studied magnetic structures via their indirect signatures, e.g., BPs, or without resolving the magnetic fields.

In a previous paper \citep[hereafter Paper 1]{2014ApJ...789....6R}, we reported on the first direct observation of the formation of an individual photospheric magnetic element as seen by \textsc{Sunrise}/IMaX. Here, we complement that work by investigating the dynamical interaction of quiet-Sun magnetic structures with the convective flows.

\begin{figure*}
\includegraphics[width=\textwidth]{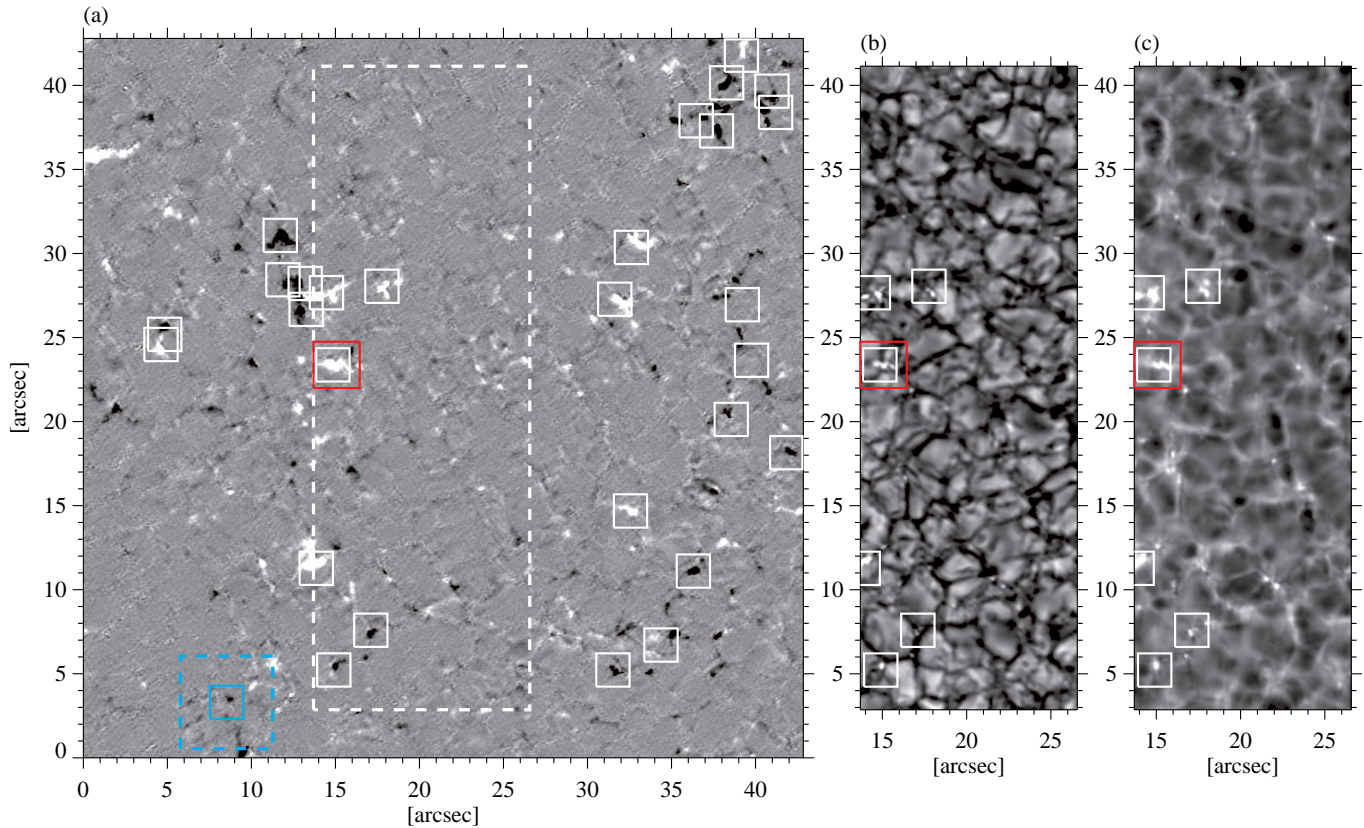}
\caption{Examples of co-spatial images from both, the IMaX and SuFI instruments. (a) IMaX Fe \textsc{I} 525.0217 nm longitudinal magnetic field covering the full FOV of about 43\arcsec\ $\times$ 43\arcsec. The longitudinal component of the magnetic field, $B\, \cos\gamma$, is linearly scaled  from -100 to 100 G. The white dashed-line rectangle, with a FOV of 13\arcsec\ $\times$ 38\arcsec,  illustrates the co-aligned area in common with the SuFI CN and Ca\,\textsc{ii}\,H images. The white boxes enclose locations where multi-cored magnetic structures are observed. The red box highlights  a feature that is examined in detail in Section \ref{MCMS}. (b) SuFI CN image. (c) SuFI Ca\,\textsc{ii}\,H image.}
\label{fig1}
\end{figure*}

\section{Observations and data reduction}

\begin{figure*}[t]
\includegraphics[width=\textwidth]{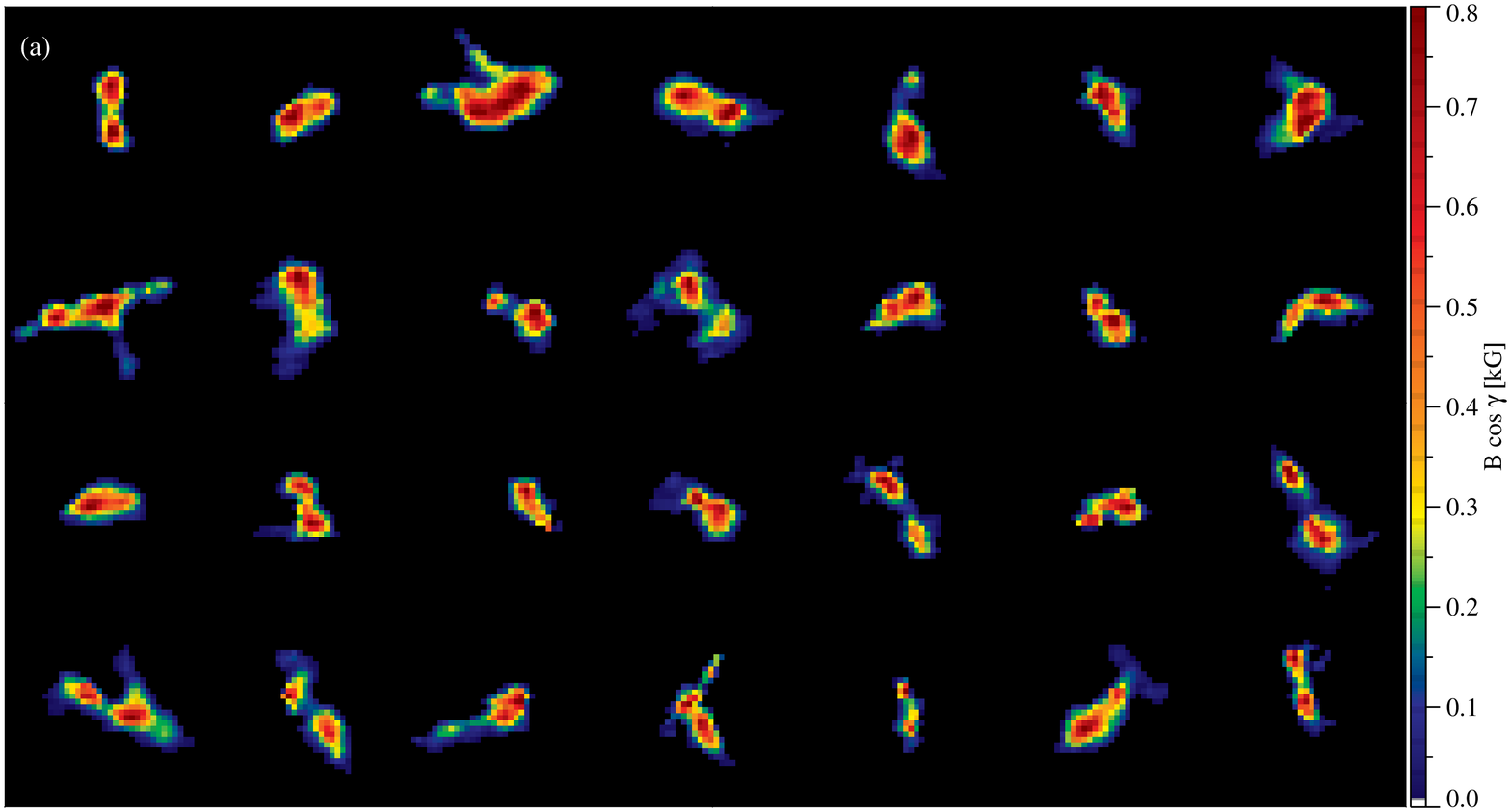}
\includegraphics[width=\textwidth]{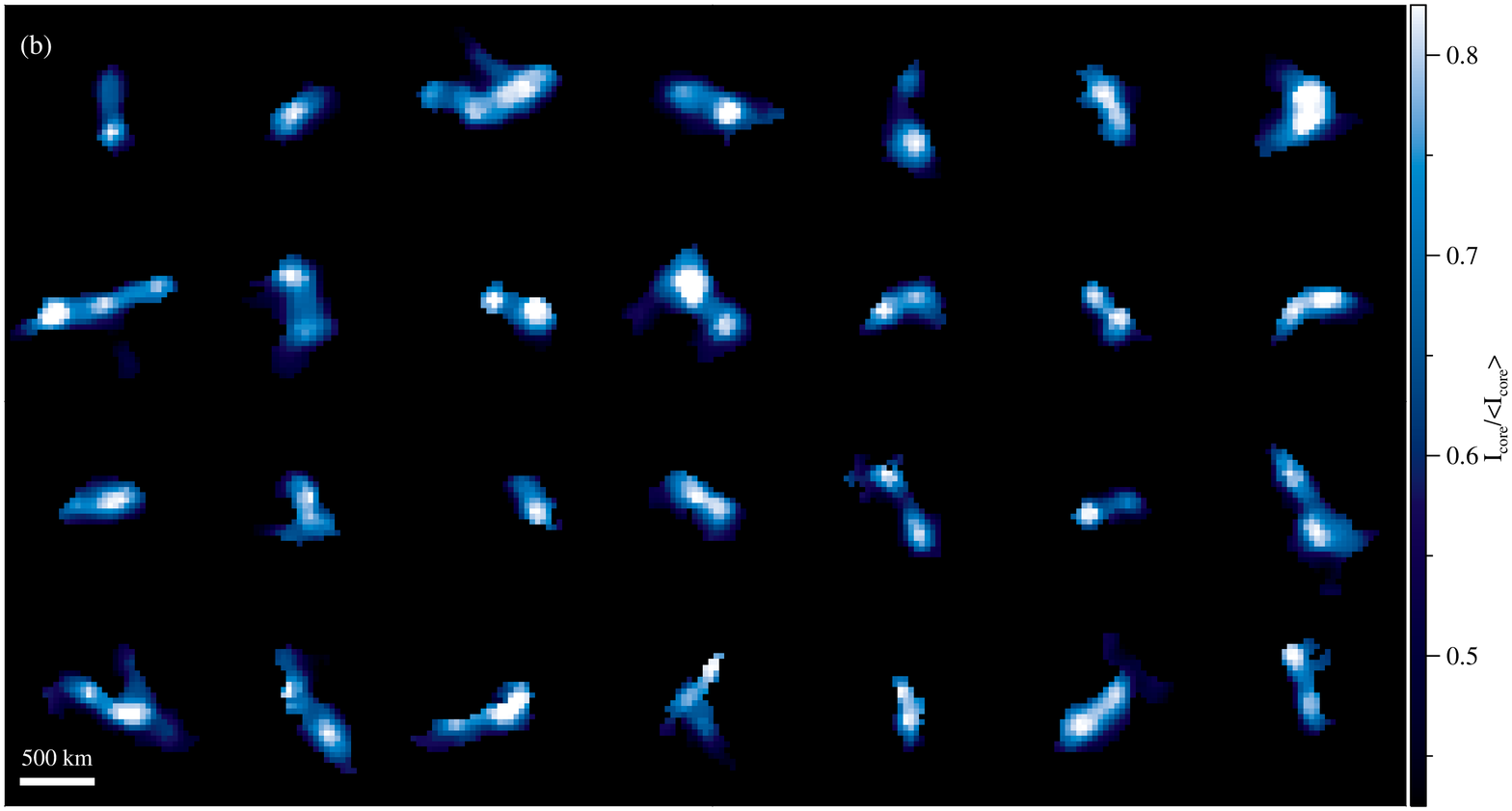}
\caption{Enlarged views of the white boxes in Figure \ref{fig1}. Each structure corresponds to a different box and is not co-temporal with the others. (a) Longitudinal component of the magnetic field. (b) Line-core intensity in units of the continuum intensity. For clarity, the plotted line-core intensity has been set to zero outside the magnetic features.}
\label{fig2}
\end{figure*}

The spectropolarimetric observational data were obtained with \textsc{Sunrise}/IMaX on 2009 June 9 from 00:36:03 UT to 00:58:46 UT, in a quiet-Sun region close to disk center. The data set of $\sim23$ minutes length has a  temporal cadence of  33 s, with a pixel size of 0\farcs 055. Throughout  the observing cycle, the full Stokes vector was sampled at five wavelength positions across the Fe \textsc{I} 525.0217 nm line (Land\'{e} factor g = 3) at $\lambda$ = -8, -4, +4, +8, and +22.7 pm from the line center \citep[V5-6 mode of IMaX; see][for details]{2011SoPh..268...57M}. For the polarization analysis, the incoming light is modulated by two liquid crystal variable retarders (LCVRs) and analyzed by a beam splitter. The spectral analysis is performed by a Fabry--P\'{e}rot interferometer based on a double-pass LiNbO$_3$ etalon.

IMaX data reduction and instrument calibrations are described by \citet{2011SoPh..268...57M}. Several procedures were used for dark-current subtraction, flat-field correction, and polarization cross-talk removal. The calibration set consisted of 30 in-focus and out-of-focus image pairs that, through phase diversity \citep{1982OptEn..21..829G,1996ApJ...466.1087P}, were used for post-facto point-spread function (PSF) retrieval. The science images were reconstructed by deconvolving this PSF from the originally recorded images. The process requires an apodization that effectively reduces the IMaX field of view (FOV) down to about 43\arcsec\ $\times$ 43\arcsec. The blueshift over the FOV produced by the Fabry--P\'{e}rot interferometer is corrected in the inferred velocity values. The instrument achieved a spectral resolution of 8.5 pm and the spatial resolution has been estimated to be 0\farcs 15--0\farcs 18 after reconstruction. The noise level in each Stokes parameter is about 3$\times$10$^{-3}$ in units of the continuum intensity, and  the rms contrast of the quiet-Sun granulation obtained from continuum data is about 13.5\% \citep{2010ApJ...723L.127S}, which testifies to the outstanding quality of IMaX images. We determine the line-core intensity by fitting the observed IMaX Stokes $I$ profiles at the sampled spectral positions by a Gaussian.

In addition to the IMaX Fe \textsc{I} 525.0217 nm images, several nearly simultaneous CN (centered at 388 nm with FWHM $\approx$ 0.8 nm) and Ca\,\textsc{ii}\,H (centered at 396.8 nm with FWHM $\approx$ 0.18 nm) filtergrams obtained with the \textsc{Sunrise} Filter Imager  \citep[SuFI;][]{2011SoPh..268...35G} are used in the present paper. The time series has a cadence of 12 s, with a pixel size of 0\farcs 0207, and a FOV of about 13\arcsec\ $\times$ 38\arcsec. The CN and Ca\,\textsc{ii}\,H bandpass images have been phase diversity reconstructed \citep{2011A&A...529A.132H}. 

Since SuFI and IMaX data have different cadences, we select those CN and  Ca\,\textsc{ii}\,H images whose observing times are closest to the IMaX observations. Note that the pixel size is also different. Thus, we increase the size of the SuFI image pixels by neighborhood averaging to a common scale with IMaX. Furthermore, we properly align the images by applying a cross-correlation technique on all simultaneous frames of Ca\,\textsc{ii}\,H and the IMaX line-core intensity, i.e., the data products having the closest BP contrast.

\section{Data analysis}

To determine the vector magnetic field and the line of sight (LOS) velocity, inversions of the full Stokes vector are carried out with the SIR code \citep{1992ApJ...398..375R} for all time steps in our series. This code numerically solves the radiative transfer equation along the LOS under the assumption of local thermodynamic equilibrium, and minimizes the difference between the measured and the computed synthetic Stokes profiles using response functions. 

Starting from the Harvard--Smithsonian Reference Atmosphere \citep{1971SoPh...18..347G} as initial guess  (with added magnetic and velocity parameter values), the temperature $T$, is modified with two nodes.\footnote{As usual in SIR, the whole atmosphere is perturbed regardless of the number of nodes. {\em Equivalent} response functions are calculated at these nodes that include the sensitivity of {\bf all} depth grid points (see \citet{1994A&A...283..129R} and \citet{2003isp..book.....D}). The number of nodes basically indicates the degree in the polynomial spline interpolation that is assumed to apply to the perturbations (not to the final stratification). In the specific case of two nodes, they are put at the first and last point of the grid ($\log \tau_{\rm c} = 1.4$ and $-4.0$) but such positions are irrelevant: the same  linear perturbation is applied independently of the node positions. However, it is worth noting that with only five wavelenght points, the temperature is not well constrained in layers above $\log\tau_{\rm c} = -2$ or below $\log\tau_{\rm c} = 0.5-0$.} The magnetic field strength $B$, the inclination and the azimuth angles $\gamma$ and $\chi$, the LOS velocity $v_{\rm LOS}$, and the microturbulent velocity $v_{\rm mic}$ are assumed to be constant with height. The magnetic filling factor $f$ is assumed to be unity and the macroturbulent velocity $v_{\rm mac}$ is set to zero due to the high spatial resolution of the data. From $B$ and $\gamma$ we also derive the longitudinal component of the magnetic field $B_{\rm long} = B\,\cos \gamma$. At each iteration step the synthetic profiles are convolved with the spectral PSF of IMaX, which was measured in the laboratory before the launch of \textsc{Sunrise} \citep{2014A&A...568A..13R}. To estimate the noise-induced uncertainty in the field strength and LOS velocity, we repeat the inversions with 100 different realizations of added noise to the observed Stokes profiles. Amplitudes of $3 \times 10^{-3}$ in units of the continuum intensity were used. The standard deviation of the 100 results is 150 G and 150 m s$^{-1}$ respectively.

Finally, we apply a p-mode subsonic filter \citep{1989ApJ...336..475T} to the continuum intensity, line-core intensity, LOS velocity, CN and Ca\,\textsc{ii}\,H images, and compute the horizontal velocity maps of the continuum intensity by means of a local correlation tracking (LCT) technique \citep{1986ApOpt..25..392N,1988ApJ...333..427N} as implemented by \citet{1994IRN..31}. Figure \ref{fig1} displays example frames of an IMaX  longitudinal magnetic field (left), a SuFI CN (center) and a Ca\,\textsc{ii}\,H (right) image after co-alignment.

\section{Multi-cored magnetic structures} \label{MCMS}

The high spatial and temporal resolution observations allow us to study the dynamics of resolved small-scale magnetic structures. This implies that we are able to track magnetic elements themselves rather than just their proxies, i.e., BPs. We use time series of $B_{\rm long}$ as context data to follow the evolution of magnetic elements. After visual inspection of each maps frame, we identify 28 groups of flux tubes evolving as single entities while the individual tubes undergo different coalescence and fragmentation processes. The locations where such magnetic structures are detected are highlighted by white boxes in Figure \ref{fig1}.

Figure \ref{fig2} shows enlarged views of these structures. The longitudinal magnetic field maps (Figure \ref{fig2}(a)) illustrate ``multi-cored'' magnetic structures that are resolvable into a series of more elemental structures, each of which might be described by a flux tube. In general, the magnetic structures are seen in the longitudinal magnetic field maps to have at least two inner cores surrounded by a common and weaker envelope. In most cases, the line core intensity maps display a  BP associated with each magnetic core (see Figure \ref{fig2}(b)). Thus, the multi-cored magnetic structures are generally characterized by groups of resolved BPs. Similar quiet-Sun bright structures have been previously observed by e.g., \citet{1996ApJ...463..365B,1998ApJ...495..973B,2004A&A...428..613B,2005A&A...435..327R} and \citet{2010ApJ...714L..31G}. Here we have been able to relate each BP group with a magnetic core group that belongs to a common underlying magnetic structure. 

The red box in Figure \ref{fig1} highlights a region of interest containing a representative example of a multi-cored magnetic structure whose dynamics we have followed. We focused on this region because it shows, in a single example, many of the processes involved in the evolution of these magnetic features. In addition, it is one of the comparatively few cases for which we have also information from SuFI. We describe it in some detail in what follows. 

\subsection{Evolution of magnetic elements} \label{ROI}

\begin{figure*}
\includegraphics[width=\textwidth]{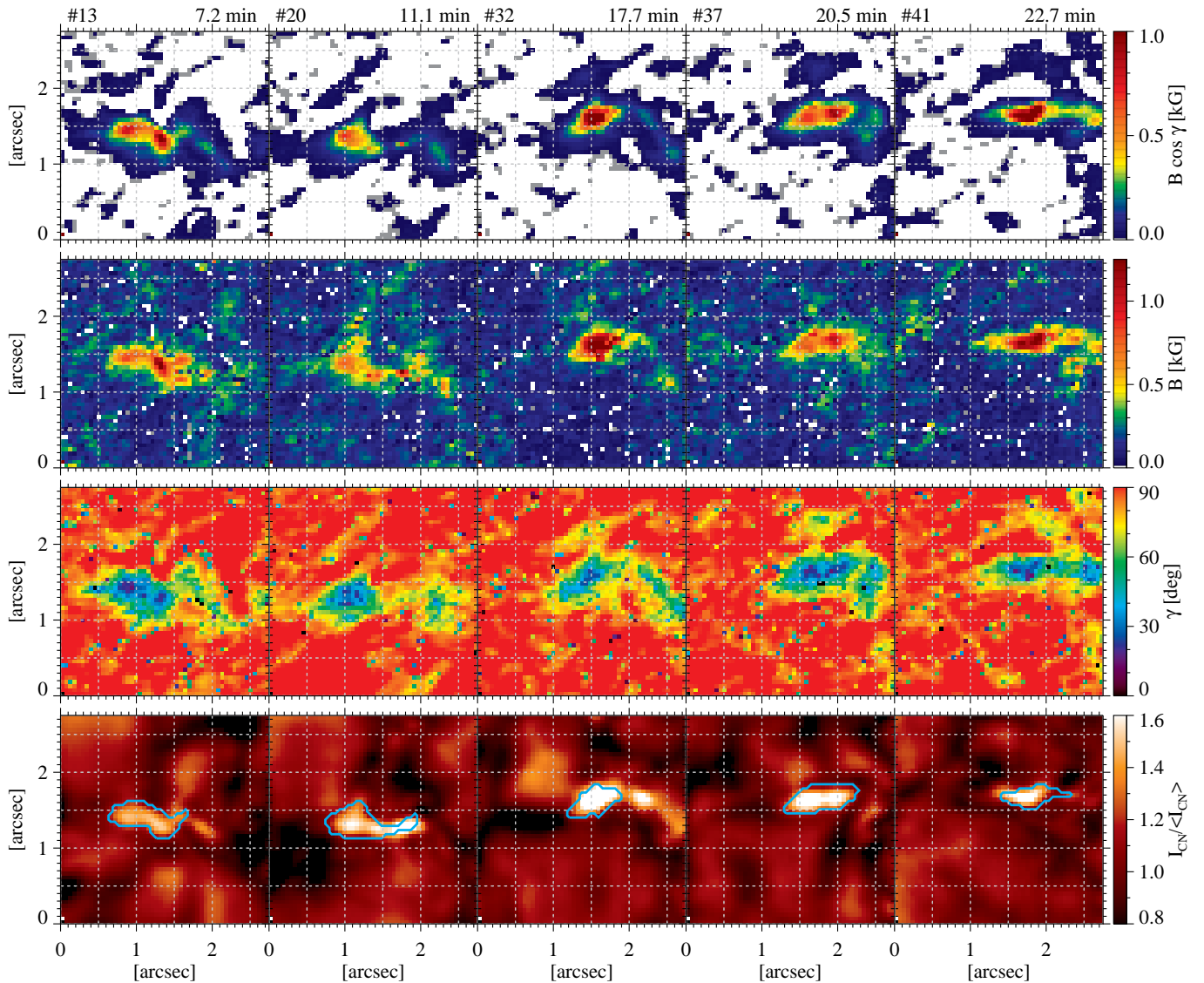}
\caption{Evolution of a multi-cored magnetic structure (red box in Figure \ref{fig1}). First row: longitudinal magnetic field. Second row: magnetic field strength $B$. Third row: magnetic field inclination $\gamma$. Fourth row: CN band images. Frame numbers (elapsed time) are given in the upper left (right) corner of each top frame.}
\label{fig3}
\end{figure*}

\begin{figure*}
\includegraphics[width=\textwidth]{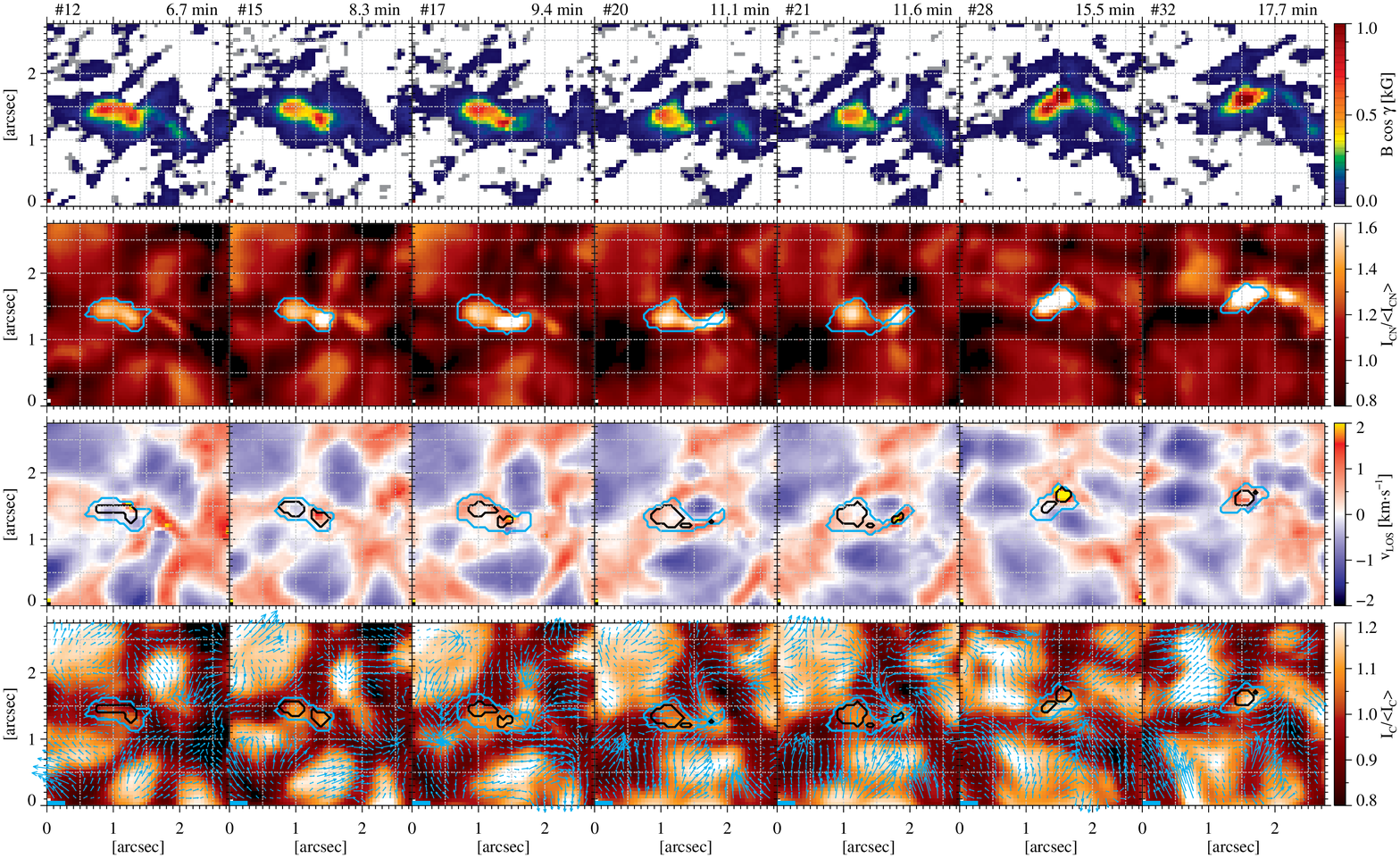}
\caption{ Close-up of the first three frames in Figure \ref{fig3} with greater temporal resolution. First row: longitudinal magnetic field $B\,\cos \gamma$. Second row: CN images. Third row: LOS velocity $v_{\rm LOS}$. Fourth row: continuum intensity. The total magnetic flux within all the black contours in a given image is constantly equal to $2.9 \times 10^{17}$ Mx.  Overplotted blue arrows in the bottom panels outline the horizontal flow field derived through the LCT technique by correlating the displayed frames with the previous ones. The length of the blue bar at coordinates [0\farcs 0, 0\farcs 0] corresponds to 1.8 km s$^{-1}$. Blue contours are the same as in Figure \ref{fig3}. This figure is also available within Animation 1 in the electronic edition of the journal.}
\label{fig4}
\end{figure*}

Figure \ref{fig3} illustrates the temporal evolution of a multi-cored magnetic structure (red box in Figure \ref{fig1}) based on five selected $B_{\rm long}$ maps (first row), magnetic field strength and inclination maps (second and third rows), and co-aligned CN maps (fourth row). The blue contour in the last row marks the periphery of the multi-cored magnetic structure. This has been selected by visual inspection in such a way that all the magnetic cores are kept within the global structure. In all frames it delineates longitudinal magnetic field iso-contours of approximately 250 G, and it encloses a magnetic flux of $(5.4\pm 1.3) \times 10^{17}$ Mx, where $\pm 1.3$ is the amount by which it changes over time. This contour will be used until the end of Section \ref{interaction}.

In frame number 13, an elongated CN bright structure, with two seemingly brighter concentrations, is observed. (The existence of two BPs can be confirmed through their evolution as seen in Animation 1.) The co-temporal longitudinal magnetic field map identifies the two brighter concentrations (BPs)  with two associated magnetic cores embedded in a more diffuse magnetic structure. These cores are also clearly observed in the field strength and inclination images. The structure is formed by two strong ($\sim1000$ G) and almost vertical ($\sim20^{\circ}$) inner cores surrounded by a common, weaker ($\sim400$ G) and more inclined ($\sim70^{\circ}$), canopy-like ring. Evidence that such rings are associated with canopies in single-cored magnetic structures has been provided by, e.g., \citet{2007A&A...476L..33R}, \citet{2012ApJ...758L..40M} and \citet{2015A&A...576A..27B}. Our highly inclined fields of the rings are fairly consistent without a-priori assumptions with the conventional picture of a canopy. However, their quantitative values may be more uncertain than those from the cores, because the polarimetric signal is weaker over the rings than over the cores. \citet{2012ApJ...758L..40M}, for instance, found smaller tube expansions with higher spectral resolution from \textsc{Sunrise}/IMaX, but no linear polarization was studied since only Stokes $I$ and $V$ were available in their case. 

The magnetic morphology suggests that the dual-core feature is formed by two magnetic elements (flux tubes) that lose their individual identity as they expand with height and merge together. Canopy merging has been observed by \citet{2015A&A...576A..27B} in a fairly different scenario: they report on individual magnetic flux concentrations whose canopies blend with those from neighbor concentrations.

The subsequent evolution shows\footnote{Better seen in Animation 1. We suggest the reader to manually play back and forth the individual frames of the  movie.} that the rightmost magnetic core splits into two (frame 20). Then the three of them merge and form an isolated magnetic element (frame 32).  Soon afterwards, however, it fragments into two apparently identical cores (frame 37) that subsequently start to fuse again into a single magnetic concentration (fourth 41). Two small remnants appear to leave the main merged core during the merging process itself. They follow an independent evolution (best seen in the CN maps) until the end of the time series through which both merge into a different (weaker) magnetic structure. Despite these recurrent fragmentation and coalescence processes, the magnetic cores keep sharing the same canopy over the whole time. These processes are analyzed in more detail in the following section.

Since we have observed that CN BPs are good proxies of magnetic cores and this is also true for those seen in the G band \citep{2001IAUS..203..287K}, the above-described evolutionary behavior is consistent with the photometric observations of, e.g., \citet{1996ApJ...463..365B}. In the light of our co-aligned spectropolarimetric observations we are in a position to assert that such BP groups, that keep together for periods much longer than a granule lifetime  \citep[up to 70 minute in][]{1998ApJ...495..973B}, can be members of the same magnetic structures.

\subsection{Interaction with granular convection} \label{interaction}

\begin{figure}
\includegraphics[scale=0.82]{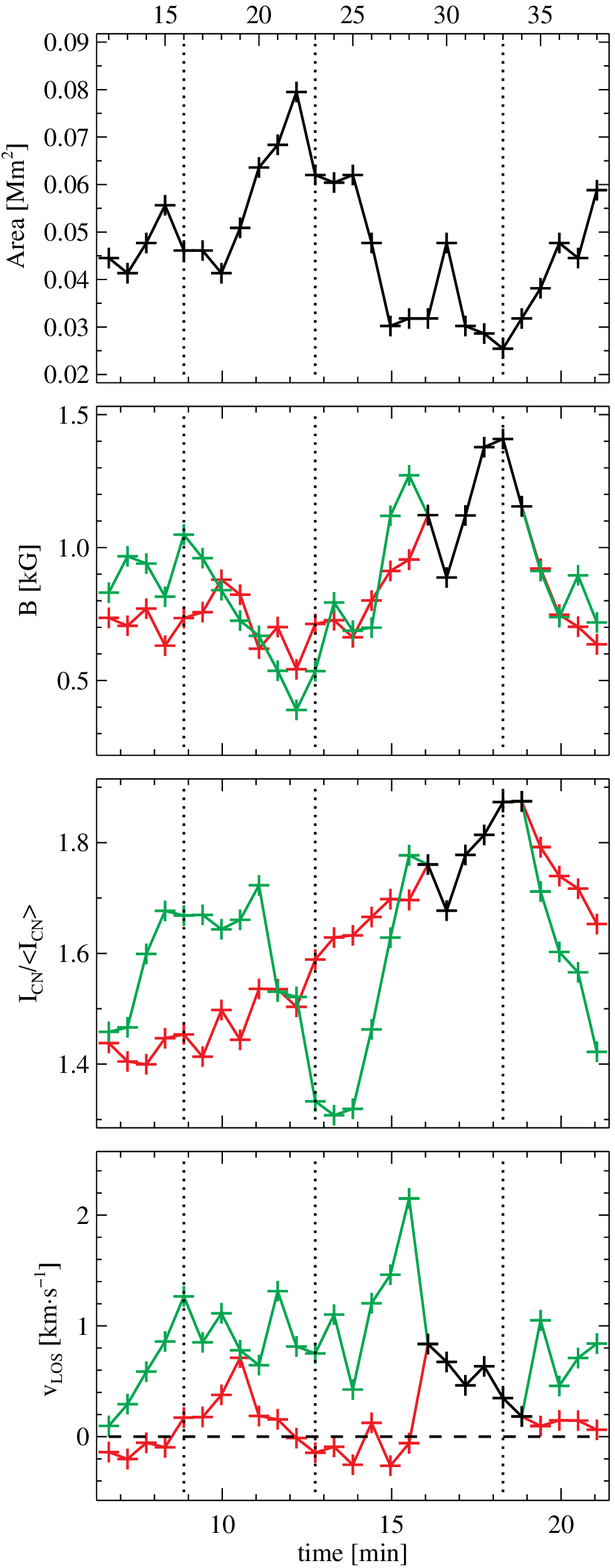}
\caption{Evolution of relevant quantities with full temporal resolution, for frames 11-38. First plot: the area within the black contour (enclosing a time-independent magnetic flux of $2.9 \times 10^{17}$ Mx). The x-axis at the top of the panel marks the frame numbers. Second plot: magnetic field strength.  Third plot: CN intensity. Fourth plot: LOS velocity. The plots display values of the corresponding quantities averaged over nine pixels centered around the centroid of the magnetic cores in the longitudinal magnetic field maps. Red (green) lines stand for the leftmost (rightmost) magnetic core and the black line is used when a single magnetic core is observed.}
\label{fig5}
\end{figure}

Figure \ref{fig4} indicates the different processes that take place during the evolution of the multi-cored magnetic structure due to its interaction with the local granular convection flows. From top to bottom, the rows show longitudinal magnetic field maps, CN intensity, LOS velocity, and continuum intensity maps. This figure is complemented by Animation 1, which is included in the electronic edition of the journal. In the animation we also display the Ca\,\textsc{ii}\,H intensity maps. As in Figure \ref{fig3}, the blue contours mark the periphery of the multi-cored magnetic structure. The new black contours have been created to follow the evolution of the individual magnetic cores and delineate a set of regions whose summed magnetic flux is constantly equal to  $2.9 \times 10^{17}$ Mx throughout the period of observation. These flux contours are constructed by starting from the most intense pixels in the longitudinal magnetic field map and then gradually expanding the contour by lowering the $B_{\rm long}$ for pixels included inside it. The black contours thus outline the magnetic cores. Finally, blue arrows show the horizontal velocity maps inferred through the LCT technique by correlating the displayed frames with the previous ones.

In Figure \ref{fig5} we quantitatively analyze the evolution of the multi-cored magnetic structure shown in Figure \ref{fig4}. For this purpose, we manually track the magnetic cores in the longitudinal magnetic field maps. The first row of panels in Figure \ref{fig5} displays the evolution of the area enclosed by our constant-flux region of  $2.9 \times 10^{17}$ Mx. This area is delimited by black contours in Figure \ref{fig4}. From top to bottom the other panels show the evolution of the field strength, CN intensity, and LOS velocity for each of the magnetic cores. To increase the signal-to-noise ratio in the magnetic core physical parameters, we represent averages over 9 pixels around their $B_{\rm long}$ centroid. The red and green lines correspond to the leftmost and rightmost cores respectively, while a black line is drawn when the two cores merge into one. The vertical dotted lines corresponds to the end of the different phases described in the following sections.

\subsubsection{Intensification by granule compression} \label{compression}

From frame 12 to 16, the leftmost magnetic core (red lines in Figure \ref{fig5}) stays at rest whereas the rightmost magnetic core is compressed between two granules (Figure \ref{fig4} and Animation 1) . The upper granule (at coordinates [1\farcs 75,2\arcsec] in frame 12) moves toward the magnetic core, while the lower granule (at coordinates [1\farcs 25,0\farcs 25]) expands. This compression process results in the intensification of the magnetic core. The field strength increases from about 800 G to about 1100 G (green line in Figure \ref{fig5}) as the CN intensity also rises.  Meanwhile, the LOS velocity grows from nearly 0 to 1.1 km s$^{-1}$ (average) with a peak velocity of up to 3 km s$^{-1}$. This maximum downflow is reached at frame 16 within the rightmost core close to a small upflow feature that emerges at the edge of the magnetic structure. As soon as this small-scale downflow/upflow feature appears, a co-spatial BP is detected in the Ca\,\textsc{ii}\,H images (see Animation 1).

Such a nearly simultaneous small-scale downward/upward velocity pattern was first observed within a magnetic element (Paper 1) and, later, close to many BPs visible in the line core of Fe \textsc{I} 525.0217 nm \citep{2014ApJ...796...79U}. In Paper 1, this pattern was detected at the end of two consecutive magnetic field intensification processes. The isolated magnetic element was compressed by all surrounding granules, and both intensification processes led to a reduction in the area of the flux concentration and an enhancement of its field strength. In our new observations, however, this phase dose not contribute much to decreasing the area of the global magnetic structure (top panel in Figure \ref{fig5}). This is manly due to the small size of the rightmost magnetic core compared to the entire area covered by magnetic flux.

\subsubsection{Fragmentation} \label{fragmentation}

After the intensification phase the rightmost magnetic core and its related CN BP get elongated (frame 17 in Figure \ref{fig4}) as a consequence of the compression.  The ``squeezing'' ends by fragmenting the magnetic core in two (frames 17-21). The squeezing is also well illustrated by the horizontal velocity arrows in the bottom panels of  Figure \ref{fig4}. For simplicity, in Figure \ref{fig5} we only show the evolution of the rightmost fragment, whose field strength and CN intensity drops abruptly (green line up to frame 23). The decrease of the field strength in the resulting fragments leads to the increase of the area enclosed in the contour of constant magnetic flux (top panel in Figure \ref{fig5}). Small variations in the average LOS velocity accompany this process. 

\subsubsection{Coalescence and further fragmentation} \label{coalescence}

\begin{figure*}
\includegraphics[width=\textwidth]{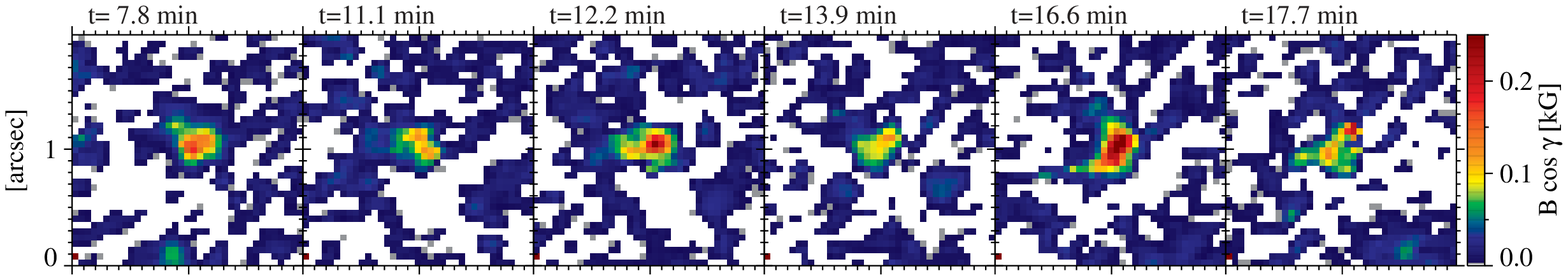}
\includegraphics[width=\textwidth]{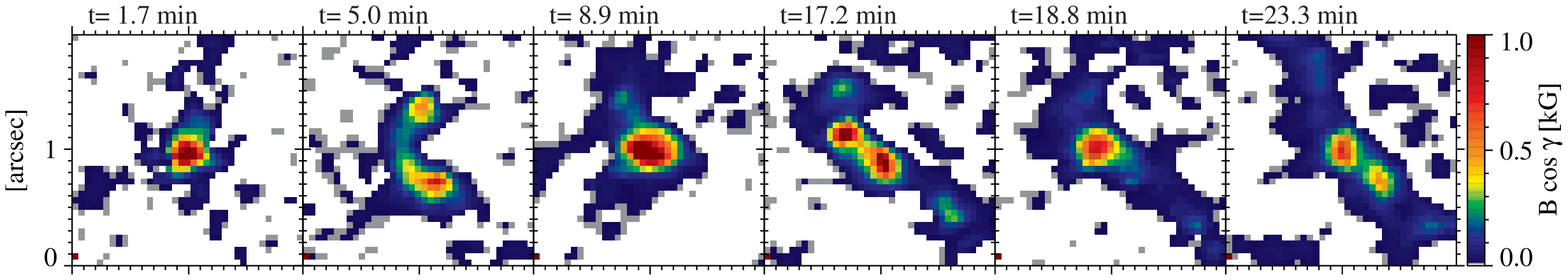}
\includegraphics[width=\textwidth]{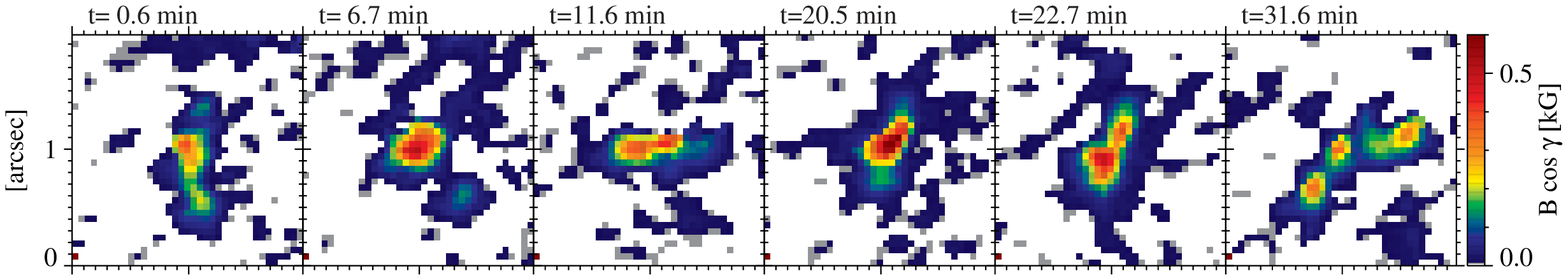}
\includegraphics[width=\textwidth]{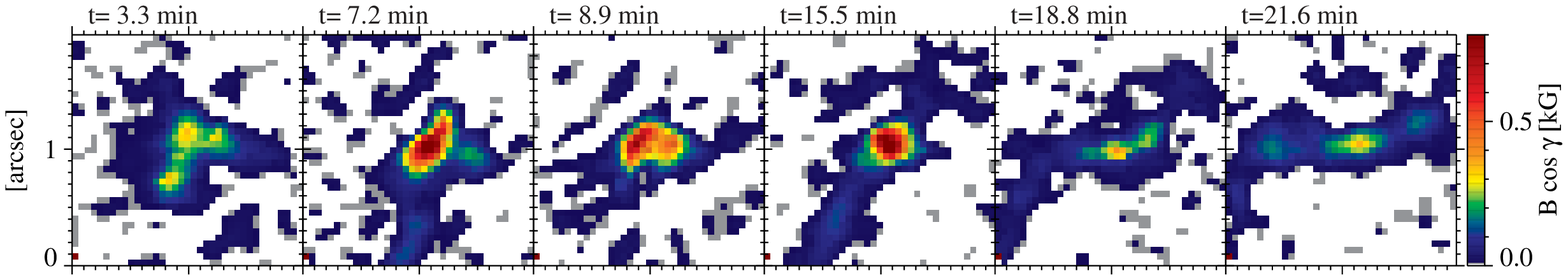}
\caption{Evolution of the longitudinal magnetic field of the four magnetic structures analyzed by \citet{2011ApJ...730L..37M}. The structure at the top row is located within the blue solid square in Figure \ref{fig1}. The other tree features corresponds to other IMaX time series and consequently their location is not shown in Figure \ref{fig1}.}
\label{fig6}
\end{figure*}

At the end of the fragmentation phase the upper granule fades away and the surrounding granules start to fill the ``empty'' space (see frame 21 to 32 in Animation 1) soon afterwards. In this way, the three magnetic cores are advected to the wide space left by the fading granule, and compressed by the surrounding granules until they merge into a single magnetic element (frame 32 in Figure \ref{fig4}). The advection of magnetic cores by the proper motions of the neighboring granules is also well illustrated by the horizontal velocity flow field. During this compression phase a strong downflow is detected within the rightmost magnetic core (frame 28) and a small upflow in its surroundings (frame 32). As soon as the downflow appears, a new co-spatial bright feature is detected in the Ca\,\textsc{ii}\,H image (Animation 1). The almost co-temporal upflow that emerges at the periphery of the magnetic structure also appears co-spatial to the Ca\,\textsc{ii}\,H BP. 

The coalescence process takes place from frame 23 to 33 (Figure \ref{fig5}). Within these 5 minutes, the magnetic fields are concentrated and, because the flux is conserved, the area decreases while the field strength increases. The magnetic field reaches a strength of up to 1.4 kG, compared with the initial $\sim$ 600 G of each magnetic core. Simultaneously, the CN intensity is also enhanced nearly in phase with the field strength. The plasma within the leftmost core is approximately at rest on average while the LOS velocity increases from 1 to 2 km s$^{-1}$ within the rightmost core. Note, however, that our 9-pixel average LOS velocity can be misleading. The apparent decrease in $v_{\rm LOS}$ for the coalesced structure results from the simultaneous presence of a downflow (in the inner core) and an upflow (at its periphery).
 
The evolution continues with a new fragmentation process. In a time interval of about 3 minutes, the magnetic element splits in two (see frame 37 in Figure \ref{fig3} or Animation 1) and the different physical quantities are almost restored to their values prior to the coalescence phase (Figure \ref{fig5}).

\subsection{Magnetic field oscillations} \label{oscillations}

\begin{figure*}
\includegraphics[width=\textwidth]{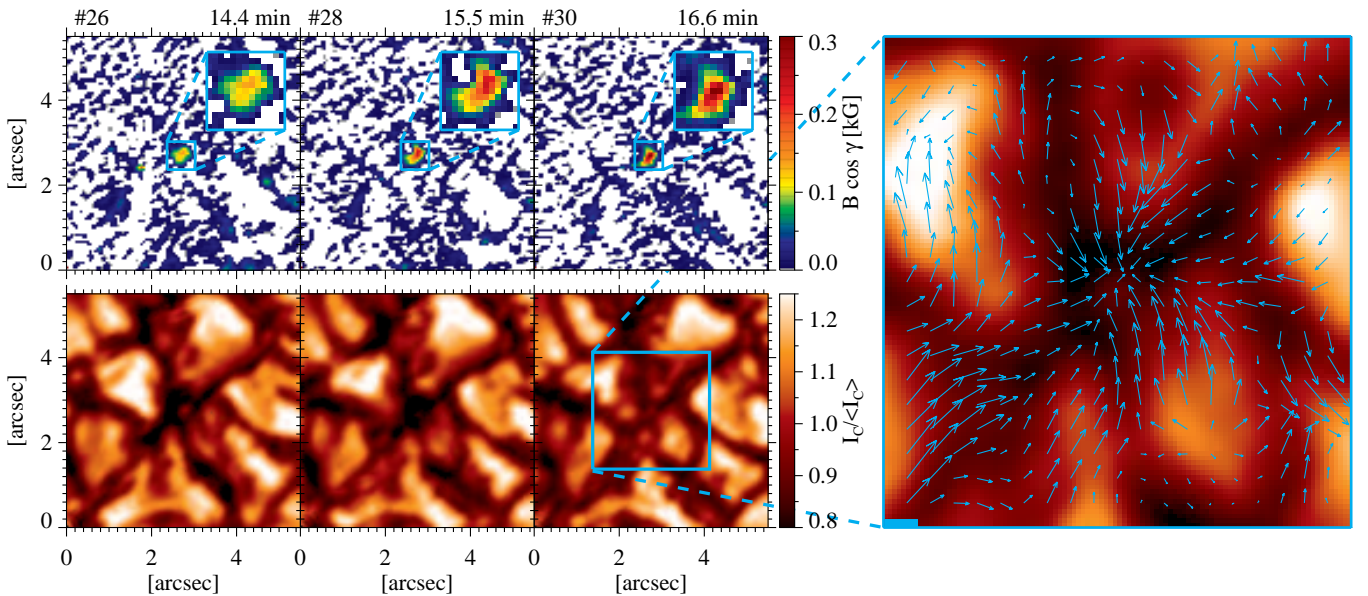}
\caption{Close-up of one of the intensification phases of the magnetic structure in the top row of Figure \ref{fig6}. The structure is located within the blue dashed square in Figure \ref{fig1}. First row: longitudinal magnetic field. Second row: continuum intensity. The rightmost panel shows the horizontal velocity maps derived through the LCT technique averaged over the intensification phase. The length of the blue bar at the lower left corner of the rightmost panel corresponds to 1.8 km s$^{-1}$.}
\label{fig7}
\end{figure*}

Driven by the local granular convective flows, the sequence of  intensification, fragmentation and coalescence events described above occur all along the evolution of the multi-cored magnetic structure. This evolutionary behavior results in oscillations of its constant magnetic flux area (top panel in Figure \ref{fig5}). Similar magnetic field oscillations where first detected in four quiet-Sun magnetic patches by \citet{2011ApJ...730L..37M} within IMaX data. They argued that the periods associated with this oscillatory pattern could be related with characteristic oscillation modes of flux tubes or, might correspond to the forcing by granular motions. Due to their compatibility with the granular lifetime and the fact that the oscillations can be strongly damped or amplified and their period abruptly modified, they favor the latter scenario. 

Here, we wonder if the oscillations found by \citet{2011ApJ...730L..37M} have also something to do with the evolution of our multi-cored magnetic structures. With the purpose of answering this question, in Figure \ref{fig6} we display  the time evolution of the longitudinal magnetic field for the four magnetic patches analyzed by them. We find that at least three of them (if not all four) are indeed multi-cored magnetic structures. The one displayed in the top row also shows hints that at some point it may be composed of at least two magnetic cores (see panels at 11.1 and 17.7 minute). However, this is not that evident as in the other three cases.

According to our new analysis, the oscillations of the three multi-cored magnetic structures can be explained through the intensification, coalescence and fragmentation processes that their inner cores suffer when they are continuously buffeted by granular flows. The damping and amplification phenomenon of oscillations and the strong changes in their periods might be caused by the changes in the number of magnetic cores contained within the structure, and by the fact that some of the fragments fade into a more diffuse magnetic field below our observational threshold (see e.g., the evolution of the magnetic feature in the second row).

It seems evident that the oscillations in at least three of the four magnetic structures are compatible with the forcing by granular motions. However, it may still be possible that oscillatory modes are present in the evolution of the magnetic structure in the first row of Figure \ref{fig6}. In order to dispel these doubts we analyze the interaction of this feature with convection. With this purpose in mind we show one of the intensification events undergone by this feature in Figure \ref{fig7}. In the first panel, the magnetic element is located in a wide space left by granules. The magnetic structure become stronger as it is compressed between the granules. During this process the surrounding granules elongate in the direction of the magnetic feature, thus forming a characteristic daisy-like pattern first described by \citet{1989SoPh..119..229M}. The compression process is well characterized by the horizontal velocity flows (rightmost panel in Figure \ref{fig7}), which point toward the magnetic feature near the center of the FOV. This shows that the oscillations correspond also in this case to the forcing by granular motions, as the magnetic feature is intensified at each of the recurrent granule compression phases.

\section{Discussion and conclusions}

We have presented direct observations of small-scale magnetic field dynamics in the quiet-Sun. This has been done with the accurate polarimetric measurements and high spatial resolution images obtained with the IMaX and SuFI instruments aboard the \textsc{Sunrise} balloon-borne stratospheric mission.
 
The results reported about several vertical magnetic cores surrounded by a common more horizontal magnetic structure suggest that we are witnessing a collection of flux concentrations in the lower photosphere that share a common canopy in the upper photospheric layers. In the photosphere, intensification, fragmentation and coalescence processes play an important role in the evolution of the individual magnetic elements. This evolution is consistent with that of their photometric counterparts (BPs) as described by \citet{1992SoPh..141...27M}, \citet{1994A&A...287..982R} and \citet{1996ApJ...463..365B}. 

The fragmentation and merging episodes appear to be governed by the evolution of the local granular convection flows.  Magnetic cores have been observed to fragment when they are ``squeezed''  or ``compressed'' by converging or expanding granules. The fragmentation of magnetic cores through the perturbation of surrounding granules may be evidence for the action of the interchange, or fluting, instability in magnetic elements. The liability to the interchange instability is indeed an inherent property of flux tubes as first noticed by \citet{1975SoPh...40..291P} and \citet{1975Ap&SS..34..347P}. The fact that the magnetic fragments share a common canopy strongly supports the theoretical predictions of \citet{1993A&A...276..236B}. His idealized model contended that magnetic slabs are liable to fluting in a limited height range around $\tau_{c} = 1$, i.e., around the solar surface. Thus, a sheet-like magnetic structure fragments into tube-like filaments. Higher up in the atmosphere, however, the single magnetic tubes lose their individual identity as they expand with height and merge into a single, stable magnetic canopy. He also conjectured that the continuous advection of the tubes back to intergranular lanes by converging granular motions might prevent further dispersion through hydrodynamic drag. 

We have also observed that soon after the splitting takes place, the resulting fragments are quickly regrouped again in intergranular lanes by the converging surrounding granules. Since the flux concentration cools the surrounding gas, it enhances the granular flows towars it \citep{1984A&A...139..435D}. This effect keeps the multi-cored magnetic structure together  during the entire 23 minute dataset. In the light of this spectropolarimetic picture, it is understandable that groups of BPs can persist for long times  \citep[up to 70 minute according to][]{1998ApJ...495..973B} while being constantly buffeted by granules.

The quantitative analysis shows that the total magnetic flux of a typical multi-cored magnetic structure remains roughly constant during its evolution. We obtain this result as we are able to spatially resolve (at least partially) this magnetic structure. We are then enabled to relate the enhancement (decrease) of the CN BP brightness during the intensification and coalescence (fragmentation) phases with the increase (decrease) of the magnetic field strength and not with changes in the local filling factor as proposed by \citet{2009ApJ...700L.145V}. It is worth noting that in contrast to them with the spatial resolution of the \textsc{Sunrise}/IMaX data ($\sim$ 0\farcs 15), we can get rid of the filling factor \citep{2010ApJ...723L.164L}. This correlation between the brightness and the field strength supports the classical picture of magnetic element radiance by the hot-wall mechanism. Accordingly, the reduced gas pressure within the flux tubes locally depresses the optical depth unity level. The less opaque magnetic flux-tube interior then causes an excess of lateral inflow of radiation into their evacuated interiors \citep{1976SoPh...50..269S,1984A&A...139..435D}, and as a consequence the magnetic elements appear brighter than their surroundings. 

In addition, as a consequence of the flux conservation, the continuous intensification, coalescence and fragmentation of magnetic cores results in oscillations of the magnetic field strength and cross-section area of the entire magnetic feature. Such oscillations were first detected in four quiet-Sun magnetic patches by \citet{2011ApJ...730L..37M}. We have found that three of them are indeed multi-cored structures, while the fourth one may have sub-resolution structure. In all these features (and in other multi-cored magnetic structures) the compression by surrounding granules plays an important role in the intensification of the magnetic field. In Paper 1 we already observed a large-amplitude variation in area and field strength within a magnetic element related to similar granule compression processes. However, due to the limited length of the observation only a single period was seen, and therefore we could not confirm that these variations were part of an oscillatory pattern. 

The excitation of the oscillations is consistent with the forcing by granular motions. The pattern we observe corresponds to the evolution of magnetic flux concentrations, whose internal structure change as they are perturbed by granular flows. Through this interaction the magnetic structures are continuously being compressed, fragmented, or their different components regrouped and hence the magnetic fields are constantly being strengthened or weakened.  

These magnetic field variations could explain the fact that brightness enhancements are observed at BPs when compressed by converging granules \citep{1992SoPh..141...27M}. They could also be the cause for the broad range of field strengths found at BPs by \citet{2007msfa.conf..165B}, p.165.

When the magnetic structure is compressed, kG field strengths are sometimes reached at the same time that strong photospheric downward motions are found within the magnetic cores. Such a correlation has been interpreted as a convective collapse by different authors \citep[e.g.,][]{2008ApJ...677L.145N,2010A&A...509A..76D}. Our findings, then, suggest that convective collapse could be triggered by granular perturbations.

The highly dynamic nature of small-scale magnetic fields found here suggests the generation of waves that could propagate up through the solar atmosphere. This is supported by the chromospheric activity that we have detected during the intensification, coalescence and fragmentation processes related with photospheric downward and upward motions. Correlation between photospheric downflows and Ca\,\textsc{ii}\,H brightenings has been explained in terms of the convective collapse process \citep{2008ApJ...680.1467S,2009A&A...504..583F}, and as disk-center photospheric traces of type II spicules \citep{2014A&A...566A.139Q}. We did not find, however, any previous mention in the literature of a relationship between  Ca\,\textsc{ii}\,H brightness and photospheric upflows as found here. In the chromosphere, high plasma velocities in the blue wing of Ca\,\textsc{ii}\,IR line have been first found by \citet{2008ApJ...679L.167L} as the disk counterpart of type II spicules. Could the  photospheric upflows that we observe here have something to do with those seen in the chromosphere? Further investigations using time series observations of comparable spatial resolution and polarimetric sensitivity at the photosphere, together with simultaneous spectroscopic information on the chromosphere, are to shed new light on these issues.

\acknowledgments

We thank M. J. Mart\'{i}nez Gonz\'{a}lez for providing the locations of magnetic structures shown in Figure \ref{fig6}. The work by I. S. R. has been funded by the Basque Government under a grant from Programa Predoctoral de Formaci\'{o}n de Personal Investigador del Departamento de Educaci\'{o}n, Universidades e Investigaci\'{o}n.
This work has been partially funded by the Spanish Ministerio de Econom\'{\i}a y Competitividad, 
through Projects No. ESP2013-47349-C6-1-R  and ESP2014-56169-C6-1-R, including a percentage from European FEDER funds.
The German contribution has been funded by the Bundesministerium f\"ur Wirtschaft und Technologie through Deutsches Zentrum f\"ur Luft- und Raumfahrt e.V. (DLR), grant number 50 OU 0401, and by the Innovationsfond of the President of the Max Planck Society (MPG). 
This work was partly supported by the BK21 plus program through the National Research Foundation (NRF) funded by the Ministry of Education of Korea.

\clearpage



\clearpage






\begin{thebibliography}{}

\bibitem[Barthol et al.(2011)]{2011SoPh..268....1B} Barthol, P., Gandorfer, 
A., Solanki, S.~K., et al.\ 2011, \solphys, 268, 1

\bibitem[Beck et al.(2007)]{2007msfa.conf..165B} Beck, C., Mikurda, K., 
Bellot Rubio, L.~R., Schlichenmaier, R., S\"{u}tterlin, P.\ 2007, Modern Solar Facilities--Advanced Solar Science, ed. F. Kneer, K. G. Puschmann \& A. D. Wittmann. Published by Universit\"{a}tsverlag G\"{o}ttingen, 165

\bibitem[Berger 
\& Title(1996)]{1996ApJ...463..365B} Berger, T.~E., \& Title, A.~M.\ 1996, \apj, 463, 365

\bibitem[Berger et al.(1998)]{1998ApJ...495..973B} Berger, T.~E., 
L{\"o}fdahl, M.~G., Shine, R.~S., \& Title, A.~M.\ 1998, \apj, 495, 973

\bibitem[Berger 
\& Title(2001)]{2001ApJ...553..449B} Berger, T.~E., \& Title, A.~M.\ 2001, \apj, 553, 449 

\bibitem[Berger et 
al.(2004)]{2004A&A...428..613B} Berger, T.~E., Rouppe van der Voort, L.~H.~M., L{\"o}fdahl, M.~G., et al.\ 2004, \aap, 428, 613 

\bibitem[Berkefeld et al.(2011)]{2011SoPh..268..103B} Berkefeld, T., 
Schmidt, W., Soltau, D., et al.\ 2011, \solphys, 268, 103 

\bibitem[Bonet et al.(2008)]{2008ApJ...687L.131B} Bonet, J.~A., 
M{\'a}rquez, I., S{\'a}nchez Almeida, J., Cabello, I., 
\& Domingo, V.\ 2008, \apjl, 687, L131 

\bibitem[Bonet et al.(2010)]{2010ApJ...723L.139B} Bonet, J.~A., 
M{\'a}rquez, I., S{\'a}nchez Almeida, J., et al.\ 2010, \apjl, 723, L139

\bibitem[Buehler et 
al.(2015)]{2015A&A...576A..27B} Buehler, D., Lagg, A., Solanki, S.~K., \& van Noort, M.\ 2015, \aap, 576, A27

\bibitem[B\"{u}nte et 
al.(1993a)]{1993A&A...268..299B} B\"{u}nte, M., Steiner, O., \& Pizzo, V.~J.\ 1993a, \aap, 268, 299

\bibitem[B\"{u}nte(1993b)]{1993A&A...276..236B} B\"{u}nte, M.\ 1993b, \aap, 276, 236

\bibitem[Danilovic et 
al.(2010)]{2010A&A...509A..76D} Danilovic, S., Sch{\"u}ssler, M., \& Solanki, S.~K.\ 2010, \aap, 509, A76 

\bibitem[Deinzer et 
al.(1984)]{1984A&A...139..435D} Deinzer, W., Hensler, G., Sch\"{u}ssler, M., \& Weisshaar, E.\ 1984, \aap, 139, 435 

\bibitem[del Toro Iniesta(2003)]{2003isp..book.....D} del Toro Iniesta, 
J.~C.\ 2003, Introduction to Spectropolarimetry (Cambridge: Cambridge Univ. 
Press)

\bibitem[Fischer et 
al.(2009)]{2009A&A...504..583F} Fischer, C.~E., de Wijn, A.~G., Centeno, R., Lites, B.~W., \& Keller, C.~U.\ 2009, \aap, 504, 583 

\bibitem[Gandorfer et al.(2011)]{2011SoPh..268...35G} Gandorfer, A., Grauf, 
B., Barthol, P., et al.\ 2011, \solphys, 268, 35 

\bibitem[Gingerich et al.(1971)]{1971SoPh...18..347G} Gingerich, O., Noyes, 
R.~W., Kalkofen, W., \& Cuny, Y.\ 1971, \solphys, 18, 347 

\bibitem[Gonsalves(1982)]{1982OptEn..21..829G} Gonsalves, R.~A.\ 1982, 
Optical Engineering, 21, 829

\bibitem[Goode et al.(2010)]{2010ApJ...714L..31G} Goode, P.~R., Yurchyshyn, 
V., Cao, W., et al.\ 2010, \apjl, 714, L31

\bibitem[Hirzberger et 
al.(2011)]{2011A&A...529A.132H} Hirzberger, J., Feller, A., Riethm{\"u}ller, T.~L., Gandorfer, A., \& Solanki, S.~K.\ 2011, \aap, 529, A132 

\bibitem[Keller(1992)]{1992Natur.359..307K} Keller, C.~U.\ 1992, \nat, 359, 
307 

\bibitem[Kiselman et al.(2001)]{2001IAUS..203..287K} Kiselman, D., Rutten, 
R.~J., 
\& Plez, B.\ 2001, in Proc. IAU Symp. 203, Recent Insights into the Physics of the Sun and Heliosphere: Highlights from SOHO and Other Space Missions, ed. P\.{a}l Brekke, Bernhard Fleck, and Joseph B. Gurman. Published by Astronomical Society of the Pacific, 287

\bibitem[Krall 
\& Trivelpiece(1973)]{1973ppp..book.....K} Krall, N.~A., \& Trivelpiece, A.~W.\ 1973, Principles of Plasma Physics (Tokyo: McGraw-Hill) 

\bibitem[Lagg et al.(2010)]{2010ApJ...723L.164L} Lagg, A., Solanki, S.~K., 
Riethm{\"u}ller, T.~L., et al.\ 2010, \apjl, 723, L164 

\bibitem[Langangen et al.(2008)]{2008ApJ...679L.167L} Langangen, {\O}., De 
Pontieu, B., Carlsson, M., et al.\ 2008, \apjl, 679, L167 

\bibitem[Lites et al.(2013)]{2013SoPh..283..579L} Lites, B.~W., Akin, 
D.~L., Card, G., et al.\ 2013, \solphys, 283, 579 

\bibitem[Mart{\'{\i}}nez Gonz{\'a}lez et al.(2011)]{2011ApJ...730L..37M} 
Mart{\'{\i}}nez Gonz{\'a}lez, M.~J., Asensio Ramos, A., Manso Sainz, R., et 
al.\ 2011, \apjl, 730, L37 

\bibitem[Mart{\'{\i}}nez Gonz{\'a}lez et al.(2012)]{2012ApJ...758L..40M} 
Mart{\'{\i}}nez Gonz{\'a}lez, M.~J., Bellot Rubio, L.~R., Solanki, S.~K., 
et al.\ 2012, \apjl, 758, LL40 

\bibitem[Mart{\'{\i}}nez Pillet et al.(2011)]{2011SoPh..268...57M} 
Mart{\'{\i}}nez Pillet, V., Del Toro Iniesta, J.~C., {\'A}lvarez-Herrero, 
A., et al.\ 2011, \solphys, 268, 57

\bibitem[Meyer et al.(1977)]{1977MNRAS.179..741M} Meyer, F., Schmidt, 
H.~U., \& Weiss, N.~O.\ 1977, \mnras, 179, 741 

\bibitem[Molowny-Horas \& Yi(1994)]{1994IRN..31} Molowny-Horas, R., \& Yi, Z. 1994, Internal Rep. 31, Institute of Theoretical
Astrophysics (Oslo: Univ. Oslo)

\bibitem[Muller(1983)]{1983SoPh...85..113M} Muller, R.\ 1983, \solphys, 85, 
113 

\bibitem[Muller 
\& Roudier(1992)]{1992SoPh..141...27M} Muller, R., \& Roudier, T.\ 1992, \solphys, 141, 27

\bibitem[Muller 
\& Roudier(1984)]{1984SoPh...94...33M} Muller, R., \& Roudier, T.\ 1984, \solphys, 94, 33 

\bibitem[Muller et 
al.(2000)]{2000A&A...359..373M} Muller, R., Dollfus, A., Montagne, M., Moity, J., \& Vigneau, J.\ 2000, \aap, 359, 373 

\bibitem[Muller et al.(1989)]{1989SoPh..119..229M} Muller, R., Hulot, 
J.~C., \& Roudier, T.\ 1989, \solphys, 119, 229 

\bibitem[Nagata et al.(2008)]{2008ApJ...677L.145N} Nagata, S., Tsuneta, S., 
Suematsu, Y., et al.\ 2008, \apjl, 677, L145

\bibitem[November(1986)]{1986ApOpt..25..392N} November, L.~J.\ 1986, \ao, 
25, 392 

\bibitem[November 
\& Simon(1988)]{1988ApJ...333..427N} November, L.~J., \& Simon, G.~W.\ 1988, \apj, 333, 427 

\bibitem[Parker(1975)]{1975SoPh...40..291P} Parker, E.~N.\ 1975, \solphys, 
40, 291 

\bibitem[Paxman et al.(1996)]{1996ApJ...466.1087P} Paxman, R.~G., Seldin, 
J.~H., Loefdahl, M.~G., Scharmer, G.~B., 
\& Keller, C.~U.\ 1996, \apj, 466, 1087

\bibitem[Piddington(1975)]{1975Ap&SS..34..347P} Piddington, J.~H.\ 1975, \apss, 34, 347

\bibitem[Quintero Noda et 
al.(2014)]{2014A&A...566A.139Q} Quintero Noda, C., Ruiz Cobo, B., \& Orozco Su{\'a}rez, D.\ 2014, \aap, 566, AA139 

\bibitem[Requerey et al.(2014)]{2014ApJ...789....6R} Requerey, I.~S., Del 
Toro Iniesta, J.~C., Bellot Rubio, L.~R., et al.\ 2014, \apj, 789, 6 

\bibitem[Rezaei et 
al.(2007)]{2007A&A...476L..33R} Rezaei, R., Steiner, O., Wedemeyer-B{\"o}hm, S., et al.\ 2007, \aap, 476, L33 

\bibitem[Riethm{\"u}ller et 
al.(2014)]{2014A&A...568A..13R} Riethm{\"u}ller, T.~L., Solanki, S.~K., Berdyugina, S.~V., et al.\ 2014, \aap, 568, AA13 

\bibitem[Roudier et 
al.(1994)]{1994A&A...287..982R} Roudier, T., Espagnet, O., Muller, R., \& Vigneau, J.\ 1994, \aap, 287, 982

\bibitem[Rouppe van der Voort et 
al.(2005)]{2005A&A...435..327R} Rouppe van der Voort, L.~H.~M., Hansteen, V.~H., Carlsson, M., et al.\ 2005, \aap, 435, 327 

\bibitem[Ruiz Cobo 
\& del Toro Iniesta(1992)]{1992ApJ...398..375R} Ruiz Cobo, B., \& del Toro Iniesta, J.~C.\ 1992, \apj, 398, 375

\bibitem[Ruiz Cobo 
\& del Toro Iniesta(1994)]{1994A&A...283..129R} Ruiz Cobo, B., \& del Toro Iniesta, J.~C.\ 1994, \aap, 283, 129

\bibitem[Scharmer et al.(1985)]{1985ApOpt..24.2558S} Scharmer, G.~B., 
Pettersson, L., Brown, D.~S., \& Rehn, J.\ 1985, \ao, 24, 2558 

\bibitem[Sch{\"u}ssler(1984)]{1984A&A...140..453S} Sch{\"u}ssler, M.\ 1984, \aap, 140, 453

\bibitem[Shimizu et al.(2008)]{2008ApJ...680.1467S} Shimizu, T., Lites, 
B.~W., Katsukawa, Y., et al.\ 2008, \apj, 680, 1467 

\bibitem[Solanki et al.(2010)]{2010ApJ...723L.127S} Solanki, S.~K., 
Barthol, P., Danilovic, S., et al.\ 2010, \apjl, 723, L127

\bibitem[Spruit(1976)]{1976SoPh...50..269S} Spruit, H.~C.\ 1976, \solphys, 
50, 269

\bibitem[Title et al.(1992)]{1992ApJ...393..782T} Title, A.~M., Topka, 
K.~P., Tarbell, T.~D., et al.\ 1992, \apj, 393, 782 

\bibitem[Title et al.(1989)]{1989ApJ...336..475T} Title, A.~M., Tarbell, 
T.~D., Topka, K.~P., et al.\ 1989, \apj, 336, 475

\bibitem[Utz et al.(2014)]{2014ApJ...796...79U} Utz, D., del Toro Iniesta, 
J.~C., Bellot Rubio, L.~R., et al.\ 2014, \apj, 796, 79 

\bibitem[Viticchi{\'e} et al.(2009)]{2009ApJ...700L.145V} Viticchi{\'e}, 
B., Del Moro, D., Berrilli, F., Bellot Rubio, L., 
\& Tritschler, A.\ 2009, \apjl, 700, L145 

\bibitem[Wang et al.(1995)]{1995SoPh..160..277W} Wang, J., Wang, H., Tang, 
F., Lee, J.~W., \& Zirin, H.\ 1995, \solphys, 160, 277

\bibitem[Yi 
\& Engvold(1993)]{1993SoPh..144....1Y} Yi, Z., \& Engvold, O.\ 1993, \solphys, 144, 1 

\end{thebibliography}
\end{document}